\def\year{2022}\relax
\documentclass[letterpaper]{article} % DO NOT CHANGE THIS
\usepackage{aaai22}  % DO NOT CHANGE THIS
\usepackage{times}  % DO NOT CHANGE THIS
\usepackage{helvet}  % DO NOT CHANGE THIS
\usepackage{courier}  % DO NOT CHANGE THIS
\usepackage[hyphens]{url}  % DO NOT CHANGE THIS
\usepackage{graphicx} % DO NOT CHANGE THIS
\usepackage{natbib}  % DO NOT CHANGE THIS AND DO NOT ADD ANY OPTIONS TO IT
\usepackage{caption} % DO NOT CHANGE THIS AND DO NOT ADD ANY OPTIONS TO IT
\DeclareCaptionStyle{ruled}{labelfont=normalfont,labelsep=colon,strut=off} % DO NOT CHANGE THIS
\usepackage{algorithm}
\usepackage{algorithmic}
\usepackage{amsmath}
\usepackage{multirow}
\usepackage{multicol}

\usepackage{newfloat}
\usepackage{listings}
\floatstyle{ruled}
\newfloat{listing}{tb}{lst}{}
\floatname{listing}{Listing}

\newcommand\blfootnote[1]{%
  \begingroup
  \renewcommand\thefootnote{}\footnote{#1}%
  \addtocounter{footnote}{-1}%
  \endgroup
}

\title{Zero-shot Audio Source Separation through Query-based Learning \\ from Weakly-labeled Data}
\author {
    Ke Chen\textsuperscript{1*}\blfootnote{The first two authors have equal contribution, and this work was performed while Ke Chen interned at Bytedance.}, 
    Xingjian Du\textsuperscript{2*},
    Bilei Zhu\textsuperscript{2},
    Zejun Ma\textsuperscript{2},
    Taylor Berg-Kirkpatrick\textsuperscript{1},
    Shlomo Dubnov\textsuperscript{1}
}
\affiliations {
    \textsuperscript{1} University of California San Diego, CA, USA\\
    \textsuperscript{2} Bytedance AI Lab, Shanghai, China\\
    \{knutchen, sdubnov, tberg\}@ucsd.edu,
    \{duxingjian.real, zhubilei, mazejun\}@bytedance.com
}

\begin{document}

\maketitle
\begin{abstract}
Deep learning techniques for separating audio into different sound sources face several challenges. Standard architectures require training separate models for different types of audio sources. Although some universal separators employ a single model to target multiple sources, they have difficulty generalizing to unseen sources. In this paper, we propose a three-component pipeline to train a universal audio source separator from a large, but weakly-labeled dataset: AudioSet. First, we propose a transformer-based sound event detection system for processing weakly-labeled training data. Second, we devise a query-based audio separation model that leverages this data for model training. Third, we design a latent embedding processor to encode queries that specify audio targets for separation, allowing for zero-shot generalization. Our approach uses a single model for source separation of multiple sound types, and relies solely on weakly-labeled data for training. In addition, the proposed audio separator can be used in a zero-shot setting, learning to separate types of audio sources that were never seen in training. To evaluate the separation performance, we test our model on MUSDB18, while training on the disjoint AudioSet. We further verify the zero-shot performance by conducting another experiment on audio source types that are held-out from training. The model achieves comparable Source-to-Distortion Ratio (SDR) performance to current supervised models in both cases.
\end{abstract}

\section{Introduction}
Audio source separation is a core task in the field of audio processing using artificial intelligence. The goal is to separate one or more individual constituent sources from a single recording of a mixed audio piece. Audio source separation can be applied in various downstream tasks such as audio extraction, audio transcription, and music and speech enhancement. Although there are many successful backbone architectures (e.g. Wave-U-Net, TasNet, D3Net \cite{wavunet, tasnet, d3net}), fundamental challenges and questions remain: How can the models be made to better generalize to multiple, or even unseen, types of audio sources when supervised training data is limited? Can large amounts of weakly-labeled data be used to increase generalization performance?

The first challenge is known as universal source separation, meaning that we only need a single model to separate as many sources as possible. Most models mentioned above require training a full set of model parameters for each target type of audio source. As a result, training these models is both time and memory intensive. There are several heuristic frameworks \cite{metatasnet} that leverage meta-learning to bypass this problem, but they have difficulty generalizing to diverse types of audio sources. In other words, these frameworks succeeded in combining several source separators into one model, but the number of sources is still limited.

One approach to overcome this challenge is to train a model with an audio separation dataset that contains a very large variety of sound sources. The more sound sources a model can see, the better it will generalize. However, the scarcity of the supervised separation datasets makes this process challenging. Most separation datasets contain only a few source types. For example, MUSDB18 \cite{musdb18} and DSD100 \cite{dsd100} contain music tracks of only four source types (vocal, drum, bass, and other) 
with a total duration of 5-10 hours. MedleyDB \cite{medleydb} contains 82 instrument classes but with a total duration of only 3 hours. There exists some large-scale datasets such as AudioSet \cite{audioset} and FUSS \cite{fuss}, but they contain only weakly-labeled data. AudioSet, for example, contains 2.1 million 10-sec audio samples with 527 sound events. However, only 5\% of recordings in Audioset have a localized event label \cite{audioset-strong}. For the remaining 95\% of recordings, the correct occurrence of each labeled sound event can be anywhere within the 10-sec sample. In order to leverage this large and diverse source of weakly-labeled data, we first need to localize the sound event in each audio sample, which is referred as an audio tagging task \cite{at-dcase}.

In this paper, as illustrated in Figure \ref{fig:model_arch}, we devise a pipeline\footnote{The official code is available in https://git.io/JDWQ5} that comprises of three components: a transformer-based sound event detection system ST-SED for performing time-localization in weakly-labeled training data, a query-based U-Net source separator to be trained from this data, and a latent source embedding processor that allows generalization to unseen types of audio sources. The ST-SED can localize the correct occurrences of sound events from weakly-labeled audio samples and encode them as latent source embeddings. The separator learns to separate out a target source from an audio mixture given a corresponding target source embedding query, which is produced by the embedding processor. Further, the embedding processor enables zero-shot generalization by forming queries for new audio source types that were unseen at training time. In the experiment, we find that our model can separate unseen types of audio sources, including musical instruments and held-out AudioSet's sound classes, effectively by achieving the SDR performance on par with existing state-of-the-art (SOTA) models. Our contributions are specified as follows:
\begin{figure}[t]
    \centering
    \includegraphics[width = 0.95\columnwidth]{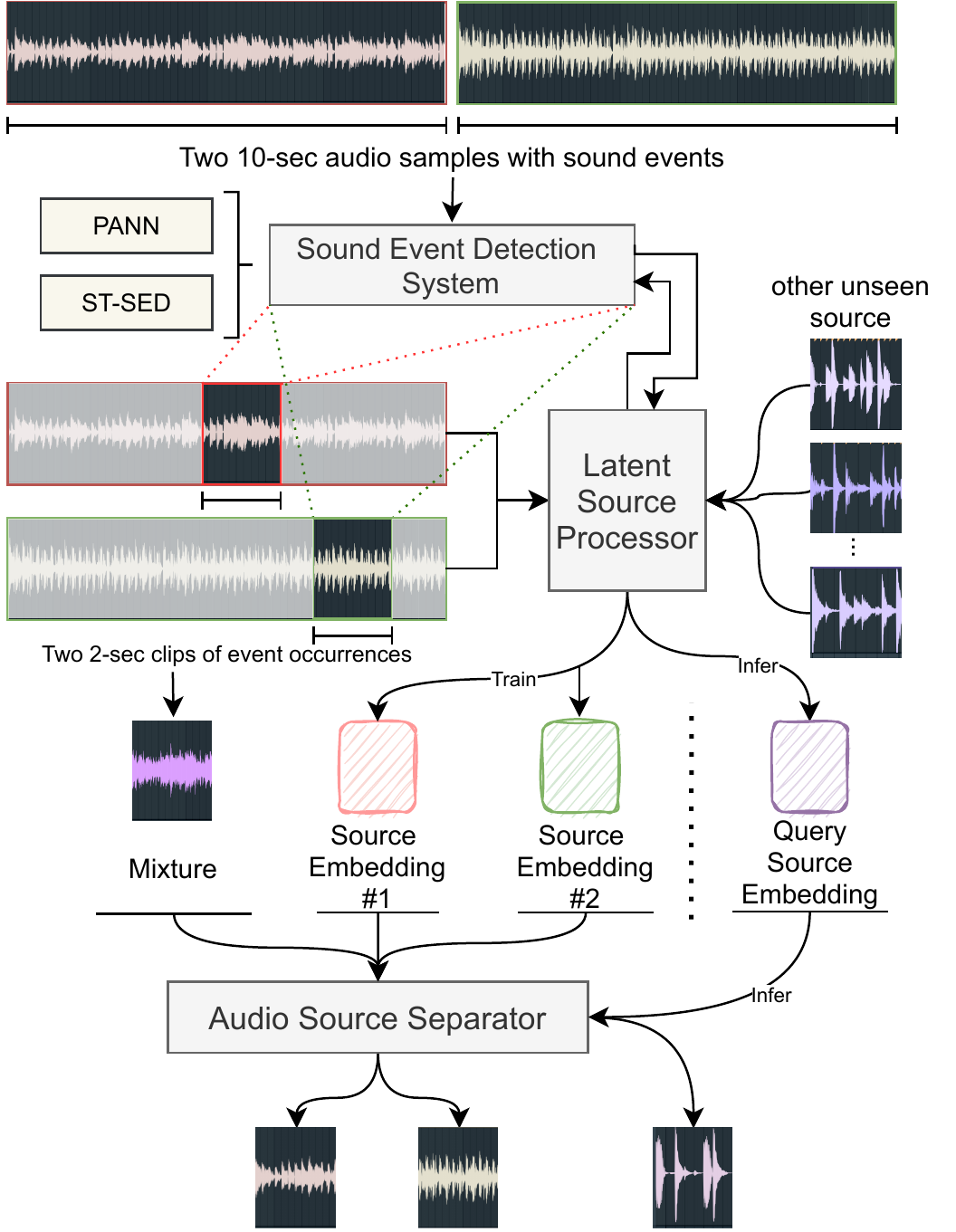}
    \caption{The architecture of our proposed zero-shot separation system. }
    \label{fig:model_arch}
\vspace{-0.5cm}
\end{figure}

\begin{itemize}
    \item We propose a complete pipeline to leverage weakly-labeled audio data in training audio source separation systems. The results show that our utilization of these data is effective.
    \item We design a transformer-based sound event detection system ST-SED. It outperforms the SOTA for sound event detection in AudioSet, while achieving a strong localization performance on the weakly-labeled data.  
    \item We employ a single latent source separator for multiple types of audio sources, which saves training time and reduces the number of parameters. Moreover, we experimentally demonstrate that our approach can support zero-shot generalization to unseen types of sources.
\end{itemize}

\section{Related Work}

\subsection{Sound Event Detection and Localization}
The sound event detection task is to classify one or more target sound events in a given audio sample. The localization task, or the audio tagging, further requires the model to output the specific time-range of events on the audio timeline. Currently, the convolutional neural network (CNN) \cite{cnn} is being widely used to detect sound events. The Pretrained Audio Neural Networks (PANN) \cite{pann} and the PSLA \cite{psla} achieve the current CNN-based SOTA for the sound event detection, with their output featuremaps serving as an empirical probability map of events within the audio timeline. For the transformer-based structure, the latest audio spectrogram transformer (AST) \cite{ast} re-purposes the visual transformer structure ViT \cite{vit} and DeiT \cite{deit} to use the transformer's class-token to predict the sound event. It achieves the best performance on the sound event detection task in AudioSet. However, it cannot directly localize the events because it outputs only a class-token instead of a featuremap. In this paper, we propose a transformer-based model ST-SED to detect and localize the sound event. Moreover, we use the ST-SED to process the weakly-labeled data that is sent downstream into the following separator. 

\begin{figure*}[t]
    \centering
    \includegraphics[width=\textwidth]{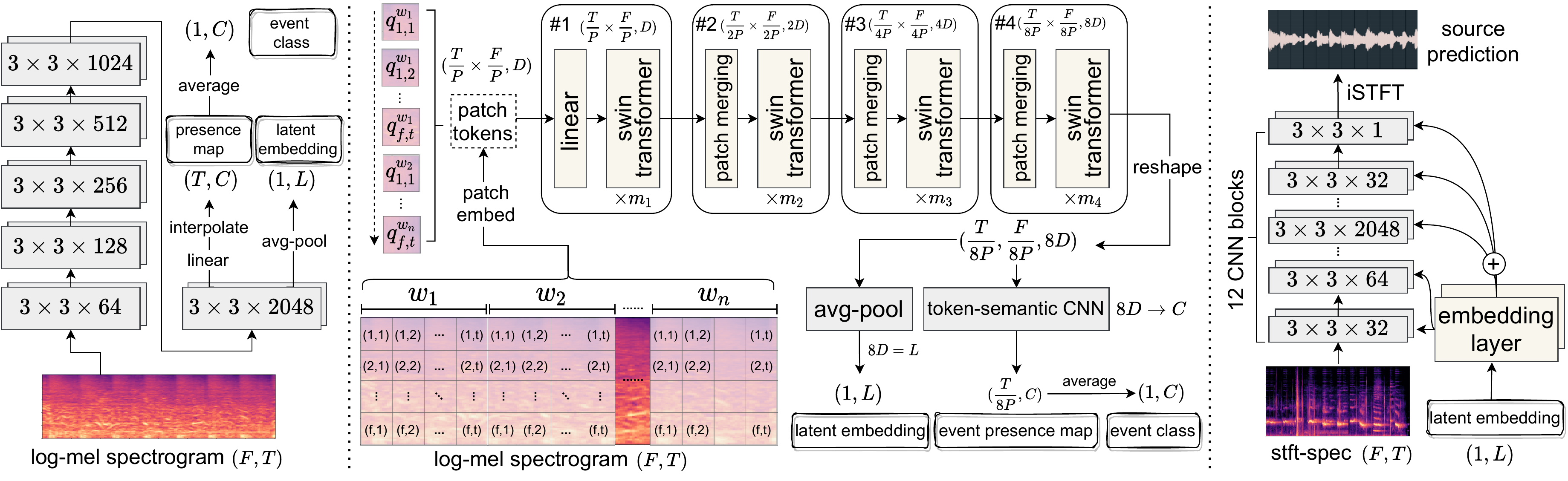}
    \caption{The network architecture of SED systems and the source separator. Left: PANN \cite{pann}; Middle: our proposed ST-SED; Right: the U-Net-based source separator. All CNNs are named as [2D-kernel size $\times$ channel size].}
    \label{fig:gsp-sed}
\vspace{-0.5cm}
\end{figure*}

\subsection{Universal Source Separation}
Universal source separation attempts to employ a single model to separate different types of sources. Currently, the query-based model AQMSP \cite{aqmsp} and the meta-learning model MetaTasNet \cite{metatasnet} can separate up to four sources in MUSDB18 dataset in the music source separation task. SuDoRM-RF \cite{sudoss}, the Uni-ConvTasNet \cite{uss}, the PANN-based separator \cite{qiangss}, and MSI-DIS \cite{zs-gus} extend the universal source separation to speech separation, environmental source separation, speech enhancement and music separation and synthesis tasks. However, most existing models require a separation dataset with clean sources and mixtures to train, and only support a limited number of sources that are seen in the training set. An ideal universal source separator should separate as many sources as possible even if they are unseen or not clearly defined in the training. In this paper, based on the architecture from \cite{qiangss}, we move further in this direction by proposing a pipeline that can use audio event samples for training a separator that generalizes to diverse and unseen sources.

\section{Methodology and Pipeline}
In this section, we introduce three components of our source separation model. The sound event detection system is established to refine the weakly-labeled data before it is used by the separation model for training. A query-based source separator is designed to separate audio into different sources. Then an embedding processor is proposed to connect the above two components and allows our model to perform separation on unseen types of audio sources.

\subsection{Sound Event Detection System}

In Audioset, each datum is a 10-sec audio sample with multiple sound events. The only accessible label is what sound events this sample contains (i.e., a multi-hot vector). However, we cannot get accurate start and end times for each sound event in a sample. This raises the problem of extracting a clip from a sample where one sound event most likely occurs (e.g., a 2-sec audio clip). As shown in the upper part of Figure \ref{fig:model_arch}, a pipeline is depicted by using a sound event detection (SED) system to process the weakly-labeled data. This system is designed to localize a 2-sec audio clip from a 10-sec sample, which will serve as an accurate sound event occurrence.

In this section, we will first briefly introduce an existing SOTA system: Pretrained Audio Neural Networks (PANN) (Left), which serves as the main model to compare in both sound event detection and localization experiments. Then we introduce our proposed system ST-SED (Middle) that leads to better performance than PANN.
% There are two models of sound event detection we use in this paper: the Pretrained Audio Neural Networks (PANN) (Left) \cite{pann}, an existing state-of-the-art system, and our proposed ST-SED (Middle). 

\subsubsection{Pretrained Audio Neural Networks}
As shown in the left of Figure \ref{fig:gsp-sed}, PANN contains VGG-like CNNs \cite{vggnet} to convert an audio mel-spectrogram into a $(T, C)$ featuremap, where $T$ is the number of time frames and $C$ is the number of sound event classes. The model averages the featuremap over the time axis to obtain a final probability vector $(1, C)$ and computes the binary cross-entropy loss between it and the groudtruth label. Since CNNs can capture the information in each time window, the featuremap $(T, C)$ is empircally regarded as a presence probability map of each sound event at each time frame. When determining the latent source embedding for the following pipeline, the penultimate layer's output $(T, L)$ can be used to obtain its averaged vector $(1, L)$ as the latent source embedding. 

\subsubsection{Swin Token-Semantic Transformer for SED}
The transformer structure \cite{transformer} and the token-semantic module \cite{tscam} have been widely used in the image classification and segmentation task and achieve better performance. In this paper, we expect to bring similar improvements to the sound event detection and audio tagging task, which then will contribute also to the separation task. As mentioned in the related work, the audio spectrogram transformer (AST) cannot be applied to audio tagging. Therefore, we refer to swin-transformer \cite{swintransformer} in order to propose a swin token-semantic transformer for sound event detection (ST-SED). In the middle of Figure \ref{fig:gsp-sed}, a mel-spectrogram is cut into different patch tokens with a patch-embed CNN and sent into the transformer in order. We make the time and frequency lengths of the patch equal as $P \times P$. 
% In that, frequency bins on the same time frame might be grouped into different patches. 
Further, to better capture the relationship between frequency bins of the same time frame, we first split the mel-spectrogram into windows $w_1, w_2, ..., w_n$ and then split the patches in each window. The order of tokens $Q$ follows \textbf{time$\to$frequency$\to$window} as:
\begin{align*}
    Q= \{&q^{w_1}_{1,1},q^{w_1}_{1,2}, ..., q^{w_1}_{1,t}, 
         q^{w_1}_{2,1}, q^{w_1}_{2,2}, ..., q^{w_1}_{2,t}, ..., q^{w_1}_{f,t}, \\
         &q^{w_2}_{1,1},q^{w_2}_{1,2}, ..., q^{w_2}_{1,t}, 
         q^{w_2}_{2,1}, q^{w_2}_{2,2}, ..., q^{w_2}_{2,t}, ..., q^{w_2}_{f,t}, \\
         &q^{w_3}_{1,1}, ..., q^{w_3}_{f,t}, q^{w_4}_{1,1}, ..., q^{w_4}_{f,t}, ..., q^{w_n}_{f,t} \}
\end{align*}
Where $t=\frac{T}{P}, f=\frac{F}{P}$, $n$ is the number of time windows, and $q^{w_k}_{i,j}$ denotes the patch in the position shown by Figure \ref{fig:gsp-sed}.
The patch tokens pass through several network groups, each of which contains several transformer-encoder blocks. Between every two groups, we apply a patch-merge layer to reduce the number of tokens to construct a hierarchical representation. Each transformer-encoder block is a swin-transformer block with the shifted window attention module \cite{swintransformer}, a modified self-attention module to improve the training efficiency. As illustrated in Figure \ref{fig:gsp-sed}, the shape of the patch tokens is reduced by 8 times from $(\frac{T}{P} \times \frac{F}{P}, D)$ to $(\frac{T}{8P} \times \frac{F}{8P}, 8D)$ after 4 network groups.

We reshape the final block's output to $(\frac{T}{8P}, \frac{F}{8P}, 8D)$. Then, we apply a token-semantic 2D-CNN \cite{tscam} with kernel size $(3, \frac{F}{8P})$ and padding size $(1,0)$ to integrate all frequency bins, meanwhile map the channel size $8D$ into the sound event classes $C$. The output $(\frac{T}{8P}, C)$ is regarded as a featuremap within time frames in a certain resolution. Finally, we average the featuremap as the final vector $(1, C)$ and compute the binary cross-entropy loss with the groundtruth label. 
Different from traditional visual transformers and AST, our proposed ST-SED does not use the class-token but the averaged final vector from the token-semantic layer to indicate the sound event. This makes the localization of sound events available in the output. In the practical scenario, we could use the featuremap $(\frac{T}{8P}, C)$ to localize sound events. And if we set $8D = L$, the averaged vector $(1, L)$ of the featuremap $(\frac{T}{8P}, L)$ can be used as the latent source embedding in line with PANN.

\subsection{Query-based Source Separator}

By SED systems, we can localize the most possible occurrence of a given sound event in an audio sample. Then, as shown in the Figure \ref{fig:model_arch}, suppose that we want to localize the sound event $s_1$ in the sample $x_1$ and another event $s_2$ in $x_2$, we feed $x_1, x_2$ into the SED system to obtain two featuremaps $m_1, m_2$. From $m_1, m_2$ we can find the time frame $t_1,t_2$ of the maximum probability on $s_1, s_2$, respectively. Finally, we could get two 2-sec clips $c_1, c_2$ as the most possible occurrences of $s_1, s_2$ by assigning $t_1, t_2$ as center frames on two clips, respectively. 

Subsequently, we resend two clips $c_1, c_2$ into the SED system to obtain two source embeddings $e_1, e_2$. Each latent source embedding $(1, L)$ is incorporated into the source separation model to specify which source needs to be separated. The incorporation mechanism will be introduced in detail in the following paragraphs.

After we collect $c_1, c_2, e_1, e_2$, we mix two clips as $c=c_1+c_2$ with energy normalization. Then we send two training triplets $(c, c_1, e_1), (c, c_2, e_2)$ into the separator $f$, respectively. We let the separator to learn the following regression:
\begin{equation}
    f(c_1 + c_2,  e_j) \mapsto c_j,j \in \{1,2\}. 
\end{equation}

\begin{figure}[t]
    \centering
    \includegraphics[width = 0.6\columnwidth]{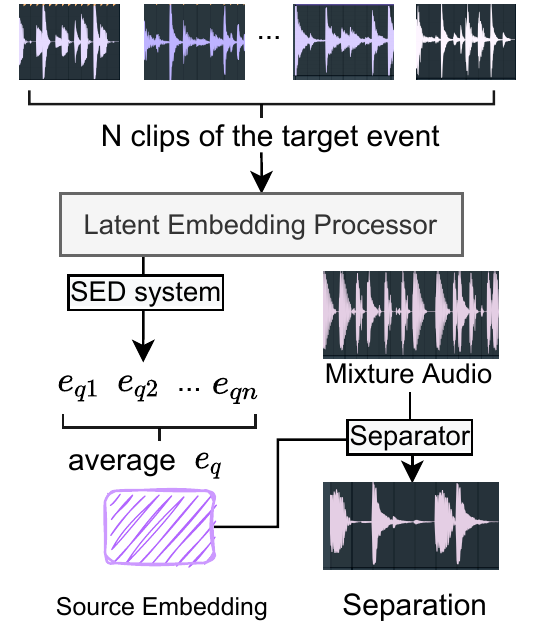}
    \caption{The mechanism to separate an audio into any given source. We collect $N$ clean clips of the target event. Then we take the average of latent source embeddings as the query embedding $e_q$. The separator receives the embedding then performs the separation on the given audio. }
    \label{fig:model_zero}
    \vspace{-0.5cm}
\end{figure}

As shown in the right of Figure \ref{fig:gsp-sed}, we base on U-Net \cite{unet} to construct our source separator, which contains a stack of downsampling and upsampling CNNs. The mixture clip $c$ is converted into the spectrogram by Short-time Fourier Transform (STFT). In each CNN block, the latent source embedding $e_j$ is incorporated by two embedding layers producing two featuremaps and added into the audio featuremaps before passing through the next block. Therefore, the network will learn the relationship between the source embedding and the mixture, and adjust its weights to adapt to the separation of different sources. The output spectrogram of the final CNN block is converted into the separate waveform $c\prime$ by inverse STFT (iSTFT). Suppose that we have $n$ training triplets $\{(c^1,c^1_j, e^1_j), (c^2,c^2_j, e^2_j), ... ,(c^n,c^n_j, e^n_j)\}$, we apply the Mean Absolute Error (MAE) to compute the loss between separate waveforms $C^\prime=\{c^{1\prime},c^{2\prime},..., c^{n\prime\}}$ and the target source clips $C_j=\{c^1_j,c^2_j,..., c^n_j\}$:
\begin{align}
    MAE(C_j, C^\prime)=\frac{1}{n}\sum_{i=0}^n \lvert c^i_j - c^{i\prime} \rvert
\end{align}

Combining these two components together, we could utilize more datasets (i.e. containing sufficient audio samples but without separation data) in the source separation task. Indeed, it also indicates that we no longer require clean sources and mixtures for the source separation task \cite{qiangss, qiangse} if we succeed in using these datasets to achieve a good performance.

\subsection{Zero-shot Learning via Latent Source Embeddings}
The third component, the embedding processor, serves as a communicator between the SED system and the source separator. As shown in Figure \ref{fig:model_arch}, during the training, the function of the latent source embedding processor is to obtain the latent source embedding $e$ of given clips $c$ from the SED system, and send the embedding into the separator. And in the inference stage, we enable the processor to utilize this model to separate more sources that are unseen or undefined in the training set.

Formally, suppose that we need to separate an audio $x_q$ according to a query source $s_q$. In order to get the latent source embedding $e_q$, we first need to collect $N$ clean clips of this source $\{c_{q1}, c_{q2}, ..., c_{qN}\}$. Then we feed them into the SED system to obtain the latent embeddings $\{e_{q1}, e_{q2}, ..., e_{qN}\}$. The $e_q$ is obtained by taking the average of them:
\begin{align}
    e_q &= \frac{1}{N}\sum_{i=1}^N e_{qi} 
\end{align}
Then, we use $e_q$ as the query for the source $s_q$ and separate $x_q$ into the target track $f(x_q, e_q)$. A visualization of this process is depicted in Figure \ref{fig:model_zero}.

The 527 classes of Audioset are ranged from ambient natural sounds to human activity sounds. Most of them are not clean sources as they contain other backgrounds and event sounds. After training our model in Audioset, we find that the model is able to achieve a good performance on separating unseen sources. According to \cite{zs-shot}, we declare that this follows a Class-Transductive Instance-Inductive (CTII) setting of zero-shot learning \cite{zs-shot} as we train the separation model by certain types of sources and use unseen queries to let the model separate unseen sources.

\section{Experiment}
There are two experimental stages for us to train a zero-shot audio source separator. First, we need to train a SED system as the first component. Then, we train an audio source separator as the second component based on the processed data from the SED system. In the following subsections, we will introduce the experiments in these two stages.  

\linespread{1.2}
\begin{table}[t]
\centering
\begin{tabular}{cc}
\hline \hline
Model                & mAP   \\ 
\hline \hline
AudioSet Baseline \shortcite{audioset}    & 0.314   \\
% TALNet. \shortcite{talnet}    & 0.362    \\
% Kong. \shortcite{kongsed}           & 0.369 \\ 
DeepRes. \shortcite{deepres}          & 0.392   \\
PANN. \shortcite{pann} & 0.434   \\
PSLA. \shortcite{psla} & 0.443   \\ 
AST. (single) w/o. pretrain \shortcite{ast} & 0.368 \\ 
AST. (single) \shortcite{ast} & 0.459 \\
\hline
768-d ST-SED   & \textbf{0.467}  \\ 
768-d ST-SED  w/o. pretrain & 0.458 \\
2048-d ST-SED  w/o. pretrain & 0.459 \\
\hline \hline
\end{tabular}
\caption{The mAP results in Audioset evaluation set.}
\label{tab:exp-sed}
\vspace{-0.5cm}
\end{table}

\subsection{Sound Event Detection}
\subsubsection{Dataset and Training Details}
We choose AudioSet to train our sound event detection system ST-SED. It is a large-scale collection of over 2 million 10-sec audio samples and labeled with sound events from a set of 527 labels. Following the same training pipeline with \cite{ast}, we use AudioSet's full-train set (2M samples) for training the ST-SED model and its evaluation set (22K samples) for evaluation. To further evaluate the localization performance, we use DESED test set \cite{desed}, which contains 692 10-sec audio samples with strong labels (time boundaries) of 2765 events in total. All labels in DESED are the subset (10 classes) of AudioSet's sound event classes. In that, we can directly map AudioSet's classes into DESED's classes. There is no overlap between AudioSet's full-train set and DESED test set. And there is no need to use DESED training set because AudioSet's full-train set contains more training data. 

For the pre-processing of audio, all samples are converted to mono as 1 channel by 32kHz sampling rate. To compute STFTs and mel-spectrograms, we use 1024 window size and 320 hop size. As a result, each frame is $\frac{320}{32000}=0.01$ sec. The number of mel-frequency bins is $F=64$. Each 10-sec sample constructs 1000 time frames and we pad them with 24 zero-frames ($T=1024$). The shape of the output featuremap is $(1024, 527)$ ($C=527$). The patch size is $4 \times 4$ and the time window is 256 frames in length. We propose two settings for the ST-SED with a latent dimension size $L$ of 768 or 2048. We adopt the 768-d model to make use of the swin-transformer ImageNet-pretrained model for achieving a potential best result. And we adopt the 2048-d model in the following separation experiment because it shares the consistent latent dimension size with PANN's. We set 4 network groups in the ST-SED, containing 2,2,6, and 2 swin-transformer blocks respectively. 

We implement the ST-SED in PyTorch\footnote{https://pytorch.org/}, train it with a batch size of 128 and the AdamW optimizer ($\beta_1$=0.9, $\beta_2$=0.999, eps=1e-8, decay=0.05) \cite{adam} in 8 NVIDIA Tesla V-100 GPUs in parallel. We adopt a warm-up schedule by setting the learning rate as 0.05, 0.1, 0.2 in the first three epochs, then the learning rate is halved every ten epochs until it returns to 0.05.

\begin{table}[t]
\centering
\begin{tabular}{c|c|c|c}
\hline
\hline
\multicolumn{4}{c}{Validation Set: AudioSet Evaluation Set} \\
\hline
\hline
Metric-SDR: dB & mixture & clean & silence \\
\hline
527-d PANN-SEP \shortcite{qiangss} & 7.38 & 8.89 & 11.00 \\
2048-d PANN-SEP & 9.42 & 13.96 & 15.89 \\
2048-d ST-SED-SEP & \textbf{10.55} & \textbf{27.83} & \textbf{16.64} \\
\hline
\hline
\end{tabular}
\caption{The SDR performance of different models with different source embeddings in the validation set. }
\label{tab:sep-valid}
\vspace{-0.5cm}
\end{table}

\begin{table*}[t]
\centering
\resizebox{\textwidth}{!}{
\begin{tabular}{c|cccccccccc|c}
\hline
\hline
Model & Alarm & Blender & Cat & Dishes & Dog & Shaver & Frying & Water & Speech & Cleaner & Average \\ \hline \hline
PANN & 34.33 & 42.35 & 36.31 & 17.60 & 35.82 & 23.81 & 9.30 & 30.58 & \textbf{69.68} & \textbf{51.01} & 35.08\\
ST-SED &  \textbf{44.66} & \textbf{52.23} & \textbf{69.98} & \textbf{27.35} & \textbf{49.93} & \textbf{43.90} & \textbf{50.22} & \textbf{42.76} & 45.11 & 41.55 & \textbf{46.77} \\ 
\hline \hline
\end{tabular}
}
\caption{The F1-score results on each class of two models in DESED test set.}
\label{tab:exp-at}
\vspace{-0.3cm}
\end{table*}

\subsubsection{AudioSet Results}
Following the standard evaluation pipeline, we use the mean average precision (mAP) to verify the classification performance on Audioset's evaluation set. In Table \ref{tab:exp-sed}, we compare the ST-SED with previous SOTAs including the latest PANN, PSLA, and AST. Among all models, PSLA, AST, and our 768-d ST-SED apply the ImageNet-pretrained models. Specifically, PSLA uses the pretrained EfficientNet \cite{effinet}; AST uses the pretrained DeiT; and 768-d ST-SED uses the pretrained swin-transformer in Swin-T/C24 setting\footnote{https://github.com/microsoft/Swin-Transformer}. We also provide the mAP result of the 768-d ST-SED without pretraining for comparison. For the 2048-d ST-SED, we train it from zero because there is no pretrained model. For the AST, we compare our model with its single model's report instead of the ensemble one to ensure the fairness of the experiment. All ST-SEDs are converged around 30-40 epochs in about 20 hours' training.

From Table \ref{tab:exp-sed}, we find that the 768-d pretrained ST-SED achieves a new mAP SOTA as 0.467 in Audioset. Moreover, our 768-d and 2048-d ST-SEDs without pretraining can also achieve the pre-SOTA mAP as 0.458 and 0.459, while the AST without pretraining could only achieve a low mAP as 0.368. This indicates that the ST-SED is not limited to the pretraining parameters of the computer vision model, and can be used more flexibly in audio tasks.

\subsubsection{DESED Results}
We conduct an experiment on DESED test set to evaluate the localization performance of PANN and the 2048-d ST-SED. We do not include AST and PSLA since AST does not directly support the event localization and the PSLA's code is not published. We use the event-based F1-score on each class as the evaluation metric, implemented by a Python library \texttt{psds\_eval}\footnote{https://github.com/audioanalytic/psds\_eval}. 

The F1-scores on all 10 classes in DESED by two models are shown in Table \ref{tab:exp-at}. We find that the 2048-d ST-SED achieves better F1-scores on 8 classes and a better average F1-score than PANN. A large increment is on the Frying class as increasing the F1-score by 40.92. However, we also notice that the F1-scores on Speech class and Cleaner class are dropped when using ST-SED, indicating that there are still some improvements for a better localization performance. 

From the above experiments, we can conclude that the ST-SED achieves the best sound event detection results and the superior results on localization performance in AudioSet and DESED. These results are sufficient for us to use \textbf{the 2048-d ST-SED model} to conduct the following separation experiments. It is better to evaluate the ST-SED on datasets. Due to the page limit, we leave these as future work.

\subsection{Audio Source Separation}
\subsubsection{Dataset and Training Details} 
We train our audio separator in AudioSet full-train set, validate it in Audioset evaluation set, and evaluate it in MUSDB18 test set as following the 6th community-based Signal Separation Evaluation Campaign (SiSEC 2018). MUSDB18 contains 150 songs with a total duration of 3.5 hours in different genres. Each song provides a mixture track and four original stems: vocal, drum, bass, and other. All SOTAs are trained with MUSDB18 training set (100 songs) and evaluated in its test set (50 songs). Different from these SOTAs, we train our model only with Audioset full-train set other than MUSDB and directly evaluate it in MUSDB18 test set.

Since Audioset is not a natural separation dataset (i.e., no mixture data), to construct the training set and the validation set, during each training step, we sample two classes from 527 classes and randomly take each sample $x_1,x_2$ from two classes in the full-train set. We implement a balanced sampler that all classes will be sampled equally during the whole training. During the validation stage, we follow the same sampling paradigm to construct 5096 audio pairs from Audioset evaluation set and fix these pairs. By setting a fixed random seed, all models will face the same training data and the validation data. 

For the model design, our SED system has two choices: PANN or ST-SED. And the separator we apply comprises 6 encoder blocks and 6 decoder blocks. In encoder blocks, the numbers of channels are namely 32, 64, 128, 256, 512, 1024. In decoder blocks, they are reversed (i.e., from 1024 to 32). There is a final convolution kernel that converts 32 channels into the output audio channel. Batch normalization \cite{bn} and ReLU non-linearity \cite{relu} are used in each block. The final output is a spectrogram, which can be converted into the final separate audio $c\prime$ by iSTFT. Similarly, we implement our separator in PyTorch and train it with the Adam optimizer ($\beta_1$=0.9, $\beta_2$=0.999, eps=1e-8, decay=0), the learning rate 0.001 and the batch size of 64 in 8 NVIDIA Tesla V-100 GPUs in parallel.

\begin{table}[t]
\centering
\resizebox{\columnwidth}{!}{
\begin{tabular}{|c|c|c|c|c|}
\hline  
\multicolumn{5}{|c|}{Standard State-of-the-art Model}                                                   \\ \hline

Metric - Median SDR                                   & vocal         & drum          & bass          & other         \\ \hline
WaveNet \shortcite{wavenet}                    & 3.25          & 4.22          & 3.21          & 2.25          \\ \hline
WK \shortcite{wk}                      & 3.76          & 4.00          & 2.94          & 2.43          \\ \hline
RGT1 \shortcite{rgt1}                                & 3.85          & 3.44          & 2.70          & 2.63          \\ \hline
Spec-U-Net \shortcite{specunet}                               & 5.74          & 4.66          & 3.67          & 3.40          \\ \hline
UHL2 \shortcite{uhl2}                                & 5.93          & 5.92          & 5.03          & \textbf{4.19}          \\ \hline
MMDenseLSTM \shortcite{mmdenselstm}                  & \textbf{6.60} & 6.41 & 5.16          & 4.15          \\ \hline
Open Unmix  \shortcite{umx}                           & 6.32          & 5.73          & 5.23 & 4.02          \\ \hline
Demucs \shortcite{demucs}                           & 6.21  & \textbf{6.50} & \textbf{6.21} & 3.80         \\ \hline \hline \hline

\multicolumn{5}{|c|}{Query-based Model w/. MUSDB18 Training}                                                                   \\ \hline 
Metric - Median SDR                                   & vocal         & drum          & bass          & other         \\ \hline
AQMSP-Mean \shortcite{aqmsp}  & 4.90          & 4.34          & 3.09          & 3.16 \\ \hline
Meta-TasNet \shortcite{metatasnet}  & 6.40 & 5.91 & 5.58 & \textbf{4.19} \\ \hline \hline \hline

\multicolumn{5}{|c|}{Zero-shot Model w/o. MUSDB18 Training}                                                                   \\ \hline
Metric - Median SDR                                   & vocal         & drum          & bass          & other         \\ \hline

527-d PANN-SEP              & 4.16   & 0.95 & -0.86 & -2.65      \\ \hline
2048-d PANN-SEP                & 6.06          & 5.00 & 3.38 & 2.86         \\ \hline
\textbf{2048-d ST-SED-SEP}      & 6.15 $\scriptstyle\pm .22$          & 5.44 $\scriptstyle\pm .32$          & 3.80 $\scriptstyle\pm .23$ & 3.05 $\scriptstyle\pm .20$ \\ \hline

\end{tabular}}
\caption{The SDR performance in MUSDB18 test set. All models are categorized into three slots. }
\label{tab:exp_cp}
\vspace{-0.5cm}
\end{table}

\begin{table*}[t]
\resizebox{\textwidth}{!}{
\begin{tabular}{cccccccccccc}
\hline
Class  & Conversation & Whispering & Clapping & Cat & Orchestra & Aircraft & Medium Engine & Pour & Scratch & Creak & Average \\ \hline
Mixture-SDR & 9.08 & 8.04 & 9.67 &  9.49 &  9.18 & 8.47 & 8.31 & 7.92 & 8.42 & 6.56 & 8.52 \\
Clean-SDR &  17.44 & 10.50 & 17.78 & 15.01 & 10.06 & 13.09 & 14.85 & 14.28 & 15.52 & 13.79 & 14.23\\
Silence-SDR & 14.05 & 13.86 & 14.45 & 17.63 &  12.08 & 11.97 & 11.56 & 12.76 & 13.95 & 13.61 & 13.59\\
\hline
\end{tabular}
}
\caption{The SDR performance of the 2048-d ST-SED-SEP in the zero-shot verification experiment. }
\label{tab:exp_zs_f}
\vspace{-0.5cm}
\end{table*}

\subsubsection{Evaluation Metrics}
We use source-to-distortion ratio (SDR) as the metric to evaluate our separator. For the validation set, we compute three SDR metrics between the prediction and the groundtruth in different separation targets:
\begin{itemize}
    \item mixture-SDR's target: $f(c_1+c_2,  e_j) \mapsto c_j$
    \item clean-SDR's target: $f(c_j,  e_j) \mapsto c_j$
    \item silence-SDR's target: $f(c_{\neg j},  e_j) \mapsto \textbf{0}$
\end{itemize}
Where the symbol $\neg j$ denotes any clip which does not share the same class with the $j$-th clip. In our setting, $\neg 1 = 2$ and $\neg 2 = 1$. The clean SDR is to verify if the model can maintain the clean source given the self latent source embedding. The silence SDR is to verify if the model can separate nothing if there is no target source in the given audio. These help us understand if the model can be generalized to more general separation scenarios only by using the mixture training. For the testing, we only compute the mixture SDR between each stem and each original song in MUSDB18 test set. Each song is divided into 1-sec clips. The song's SDR is the median SDR over all clips. And the final SDR is the median SDR over all songs.

\subsubsection{The Choice of Source Embeddings}
We choose three source embeddings for our separator: (1) the 527-d presence probability vector from PANN, referring to \cite{qiangss}; (2) the 2048-d latent embedding from PANN's penultimate layer; and (3) the 2048-d latent embedding from ST-SED. This helps to verify if the latent source embedding can perform a better representation for separation, and if the embedding from ST-SED is better than that from PANN.

In the training and validation stage, we get each latent source embedding directly from each 2-sec clip according to the pipeline in Figure \ref{fig:model_arch}. After picking the best model in the validation set, we follow Figure \ref{fig:model_zero} to get the query source embeddings in MUSDB18. Specifically, we collect all separate tracks in the highlight version of MUSDB10 training set (30 secs in each song, 100 songs in total) and take the average of their embeddings on each source as four queries: vocal, drum, bass, and other.

\subsubsection{Separation Results}

Table \ref{tab:sep-valid} shows the SDRs of two models in the validation set. We could clearly figure out that when using the 2048-d latent source embedding, PANN achieves better performance in increasing three types of SDR by 2-4 dB than that of 527-d model. A potential reason is that the extra capacity of the 2048-d embedding space helped the model better capture the feature of the sound comparing to the 527-d probability embedding. In that, the model can receive more discriminative embeddings and perform a more accurate separation. 

Then we pick the best models of 527-d PANN-SEP, 2048-d PANN-SEP, 2048-d ST-SED-SEP and evaluate them in MUSDB18. As shown in Table \ref{tab:exp_cp}, there are three categories of models: (1) Standard Model: these models can only separate one source, in that they need to train 4 models to separate each source in MUSDB18. (2) Query-based Model: these models can separate four sources in one model. Both models in (1) and (2) require the training data in MUSDB training set and cannot generalize to separate other sources. And (3) Zero-shot Model: our proposed models can separate four sources in one model without any MUSDB18 training data. Additionally, they can even separate more sources. Specifically, for our proposed 2048-d ST-SED model, we repeat the training three times with different random seeds.

From Table \ref{tab:exp_cp} our proposed model 2048-d ST-SED-SEP outperforms PANN-SEP models in all SDRs (6.15, 5.44, 3.80, 3.05). The SDRs in vocal, drum, and bass are compatible with standard and query-based SOTAs. However, we observe a relatively low SDR in the "other" source. One possible reason is that we compute a wrong "other" embedding for separation by averaging over all source embedding of "other" in MUSDB18 training set. But this "other" embedding might not be a general embedding because “other" denotes different instruments and timbres in different tracks. Another observation lies in the relatively large standard deviations of all four instruments. One possible reason is that the separation quality is related to the random combination of training data, and different orders may cause differences on some specific types of sounds. One further improving idea is to increase the numbers of combinations (e.g., three instead of two). These sub-topics can be further researched in the future.

In summary, the most novel and surprising observation is that our proposed audio separator succeeds in separating 4 sources in MUSDB18 test set without any of its training data but only Audioset. The model performs as a zero-shot separator by using any latent source embedding collected from accessible data, to separator any source it faces.

\subsection{Zero-Shot Verification}
In this section, we conduct another experiment to let the model separate sources that are held-out from training. We first select 10 sound event classes in Audioset. Then during the training, we remove all data of these 10 classes. The model only learns how to separate clips mixed by the left 517 classes. During the evaluation, we construct 1000 ($100 \times 10$) mixture samples in Audioset evaluation set whose constituents only belong to these 10 classes. Then we calculate the mixture SDR, the clean SDR, and the silence SDR of them. 

Table \ref{tab:exp_zs_f} shows the results by the 2048-d ST-SED model. We can find that the model can still separate the held-out sources well by achieving the average mixture SDR, clean SDR, and silence SDR as 8.52 dB, 14.23 dB, and 13.59 dB. The detailed SDR distribution of these 1000 samples is depicted in the open source repository. The intrinsic reason for this good performance is that the SED system captures many features of 517 sound classes in its latent space. And it generalizes to regions of the embedding space it never saw during training, which the unseen 10 classes lie in. Finally, the separator utilizes these features in the embedding to separate the target source. The zero-shot setting of our model is essentially built by a solid feature extraction mechanism and a latent source separator.

\section{Conclusion and Future Work}\label{sec:conclusion}
In this paper, we propose a zero-shot audio source separator that can utilize weakly-labeled data to train, target different sources to separate, and support more unseen sources. We train our model in Audioset while evaluating it in MUSDB18 test set. The experimental results show that our model outperforms the query-based SOTAs, meanwhile achieves a compatible result with standard supervised models. We further verify our model in a complete zero-shot setting to prove its generalization ability. With our model, more weakly-labeled audio data can be trained for the source separation problem. And more sources can be separated via one model. In future work, since audio embeddings have been widely used in other audio tasks such as music recommendation \cite{ke-recom} and music generation \shortcite{ke-musamg, ke-cmg, ke-thesis, haowen-muspy, ke-sketchnet}, we expect to use these audio embeddings as source queries to see if they can capture different audio features and lead to better separation performance.

\bibliography{aaai22}

% 
% mixture sdr 
% average:
% 9.082823945489247 8.046578060349 9.67783843513899 9.49874189722158 9.181770348811265 8.471515088743379 8.311729768761474 7.92789694295822 8.420337112373765 6.565572826141397
% median:
% 9.024891908042854 7.955378607571102 9.126542697419634 9.505769161395012 8.308339611156612 7.8264255820810344 7.839768566504759 7.594579992144533 8.230572227696918 7.1785577825919225

% clean_sdr
% average:
% 17.445532874546743 10.506624916219792 17.789668791206303 15.014645641132919 10.068488996768735 13.0912047994443 14.851723403118436 14.287040872236755 15.522057998750421 13.798063783592136
% median:
% 16.188742108185238 10.102356025541882 16.73384435509771 14.219655082516343 8.911799709984788 10.938880220882917 13.90164853998128 12.679692480458948 14.66557993233813 12.452856621727824

% slience sdr
% average:
% 14.056143258687989 13.867407953586342 14.457495685835692 17.63133877325378 12.081137248635185 11.971941875277961 11.569751217390415 12.764191317933511 13.95559753280354 13.611155498728618
% median:
% 13.011934748686102 13.499034744874717 12.721366124325206 16.754482500576376 11.255276759621184 11.602879663311924 11.099494079075662 11.882368688039815 11.786595256058261 11.509900621632159
\end{document}